\documentclass[aps,prd,twocolumn,groupedaddress,nofootinbib,showpacs, showkeys,amssymb,10pt]{revtex4}
\usepackage[a4paper,left=25mm,right=25mm,top=25mm,bottom=25mm]{geometry}
\usepackage{latexsym,float}
\usepackage{epsfig,graphics}
\usepackage{epstopdf}
\usepackage{amssymb}
\usepackage{amsmath}
\usepackage{verbatim}
\usepackage{multirow}
\usepackage{color}
\usepackage{tikz}
\usepackage[utf8]{inputenc}
\usepackage{commath}
\usepackage{braket}
\usepackage{amssymb}
\usepackage{lineno}
\usepackage{graphicx}

\providecommand{\abs}[1]{\lvert#1\rvert}
\def\gtwid{\mathrel{\raise.3ex\hbox{$>$\kern-.75em\lower1ex\hbox{$\sim$}}}}
\def\ltwid{\mathrel{\raise.3ex\hbox{$<$\kern-.75em\lower1ex\hbox{$\sim$}}}}
\def\square{\kern1pt\vbox{\hrule height 1.2pt\hbox{\vrule width 1.2pt\hskip 3pt
			\vbox{\vskip 6pt}\hskip 3pt\vrule width 0.6pt}\hrule height 0.6pt}\kern1pt}
\begin{document}

\title{Quantum Information metric for time-dependent quantum systems and higher-order corrections }
%and its application to time-driving Quantum Phase Transitions}

\author{Davood  Momeni}
\email{davood@squ.edu.om}

\affiliation{Center for Space Research, North-West University, Mafikeng, South Africa }
\affiliation{Tomsk State Pedagogical University, TSPU, 634061 Tomsk, Russia
}
\affiliation{Department of Physics, College of Science, Sultan Qaboos University,
	\\P.O. Box 36, P.C. 123, Al-Khodh, Muscat, Sultanate of Oman}

\author{Phongpichit Channuie}
\email{channuie@gmail.com}
\affiliation{School of Science, Walailak University, Thasala, \\Nakhon Si Thammarat, 80160, Thailand}

\author{Mudhahir Al Ajmi}
\email{mudhahir@squ.edu.om}
\affiliation{Department of Physics, College of Science,\\ Sultan Qaboos University,
\\P.O. Box 36, P.C. 123, Al-Khodh, Muscat, Sultanate of Oman}

\begin{abstract}
It is well established that quantum criticality is one of the most intriguing phenomena which signals the presence of new states of matter. Without prior knowledge of the local order parameter, the quantum information metric (or fidelity susceptibility) can indicate the presence of a phase transition as well as it measures distance between quantum states. In this work, we calculate the distance between quantum states  which is equal to the fidelity susceptibility in quantum model for a time-dependent system describing a two-level atom coupled to a time-driven external field. As inspired by the Landau-Zener quantum model, we find in the present work information metric induced by fidelity susceptibility. We, for the first time, derive a higher-order rank-3 tensor as a third-order fidelity susceptibility. Having computed quantum noise function in this simple time-dependent model we show that the noise function eternally lasts long in our model. 
\end{abstract}

\date{\today}

\maketitle
	
% * <channuie@gmail.com> 2017-12-28T10:00:47.820Z:
%
% ^.

\section{Introduction}
Quantum criticality is one of the most intriguing phenomena which is crucial for interpreting a wide variety of experiments.  is well known, it signals the presence of new states of matter \cite{QPT2011}. In order to observe exotic features at quantum critical point, one has to study systems in the thermodynamic regime involving large numbers of interacting particles, which encounter experimental and theoretical limitations \cite{Cardy1996}. Despite consisting only of a single-mode cavity field and a two-level atom, the authors of Ref.\cite{RabiQPT2015} show that the Rabi system exhibits a quantum phase transition (QPT). They demonstrate that the superradiant QPT primarily studied for systems of many atoms can be achieved with systems of a single one.

In recent years, there was a great deal of interest in studying QPTs from different perspectives of quantum information science \cite{QI2000}, e.g., quantum entanglement \cite{entanglementQPT2002,entanglementQPT2008} and quantum fidelity \cite{Wen-Long You,Sun2006,Zanardi2006,GuReview}. At the phase transition point, physical observables exhibit sigular behavior governing the most dramatic manifestations of the laws of statistical and quantum mechanics. In order to probe the phase transition, the fidelity susceptibility draws one of the most promising machines in which no prior knowledge of the order parameter and the symmetry of the system are required \cite{Venuti2007}-\cite{Maity:2015rfa}. Regarding these works, the connection  between the quantum information theory and condensed matter physics can be in principle achieved which might allow us to deepen our understanding in the various condensed matter phenomena. Notice that the concept of the fidelity susceptibility was originally introduced in Ref.\cite{Wen-Long You}.
 In a recent study \cite{A. Dutta},the fidelity both in the susceptibility limit and the thermodynamic limit has
been nicely summarize. Furthermore in Ref. \cite{Mukherjee-Dutta},quantum inforamtion metric has been investigated near critical points.

An obvious physical example of QPTs using the quantum fidelity approach recently is given in \cite{GuReview},\cite{Mukherjee-Dutta}. It was illustrated that at two sides of the critical point $g_c$ of a quantum many body system the ground state wavefunctions have different structures Ref.\cite{Wei}. Then, consequently, this may lead to the overlap of the two ground states which are separated by a small distance $\delta g$ in the parameter space and then might emerge. In general, at the critical point $g_c$ the distance can be parameterized via $|\braket{\Psi_0(g)|\Psi_0(g+\delta g)}|$ which is minimum. Therefore, the structure of the ground state of a quantum many-body system experiences a significant change because the system is driven across the transition point adiabatically. As a consequence, we expect that the fidelity susceptibility should be  maximum (or even diverse) at the transition point, \cite{Wen-Long You}. We notice that in various systems many authors have investigated the QPTs from the fidelity point of view \cite{GuReview,Wen-Long You,Zanardi2007}-\cite{fs2017}. They have shown that the fidelity susceptibility can be considered to be a simple approach in determining the universality of quantum phase transitions \cite{Zanardi2007}-\cite{fs2017}. Interestingly, in \cite{MIyaji:2015mia}, the quantum information metric gravity dual in conformal field theories has just been examined.

In this work, we study the fidelity susceptibility in quantum model for a time-dependent system describing a two-level atom coupled to a time-driven external field. 
%Having only two constituent particles, the Rabi model is far
%from being in the thermodynamic limit where a QPT typically occurs; however, a ratio of the atomic transition frequency to the cavity field frequency that approaches infinity, can play the role of a thermodynamic
We analytically investigate the behavior of fidelity susceptibility in the time driven quantum model when the potential $V$ is time-dependent.
%%%%%%%%%%%%%%%%%%%%%%%%%%%%%%%%%%
The organization of the paper is as follows.  In Sec.\ref{III}, we explore the mathematical foundations for fidelity susceptibility in time-dependent systems. In Sec.\ref{IV}, the two level Landau-Zener problem  is analyzed. In Sec.\ref{V}, an experimental method based on noise function is proposed. In Sec.\ref{VI}, the higher-order correction to fidelity susceptibility is calculated. Finally, we conclude our findings in the last section.

\section{Mathematical formulation of fidelity susceptibility in time-dependent driving systems }
\label{III}
%%%%%%%%%%%%%%%%%%%%%%%%%%%%%%%
In this section we will formulate fidelity susceptibility for a general time-deriving system with two levels. Let us consider a physical system with non-perturbative time dependent Hamiltonian $H_0$ in operator form:
\begin{equation}
i \hbar\frac{\partial}{\partial t} \psi_k^{(0)}= {H}_0 \psi_k^{(0)}.
\end{equation}
Our aim is to find perturbed wavefunctions with Hamiltonian ${H}=H_0+V(t)$ when $|{V}(t)| \ll |{H}_0|$. Note that here $V$ is considered to have off diagonal components, i.e, $V_{m\neq n}=\braket{\psi_m^{(0)}|V|\psi_{n}^{(0)}}\neq0$. 
Suppose that the perturbative solution for ${H}$ can be technically written in the following form:
\begin{equation}
\Psi=\sum_k a_k \psi_k^{(0)},\label{sol2}
\end{equation}
where $a_k=a_k(t)$. Substituting (\ref{sol2}) into Schr{\"o}dinger equation and multiplying by $\psi_m^{(0)}$, we obtain:
\begin{equation}
i \hbar\frac{d a_m}{d t}= \sum_k V_{mk} (t) a_k,
\end{equation}
where
\begin{equation}
V_{mk}(t)=\int{\psi_m^{\ast(0)}}\hat{V}\psi_k^{(0)}dt
=V_{mk} e^{i\frac{E_m^{(0)}-E_k^{(0)}}{\hbar}t}.
\end{equation}
Using iteration method up to the first order, i.e. $a_k^{(0)}+a_k^{(1)}$ where $a_k^{(0)}=a_k(t=0)$, we can find the ordinary differential equation for the first-order perturbation, 
\begin{equation}
i \hbar\frac{d a_k^{(1)}}{d t}= V_{kn} (t).
\end{equation}
Finally, up to the first order perturbation theory, the total wave function is written as
\begin{equation}
\Psi_n=\sum_k a_{kn}(t) \psi_k^{(0)}.
\end{equation}
Performing an integration, we obtain
\begin{equation}
a_{kn}^{(1)}=-\frac{i}{\hbar}\int{V_{kn}(t)dt}=-\frac{i}{\hbar}\int{V_{kn}e^{iw_{kn}t}}dt.
\end{equation}
In this case, to figure out how $\chi_F$ looks like, we need the ground state wavefunction to be, 
\begin{eqnarray}
\psi_n
&=&\psi_n^{(0)}+\sum_k a_{kn}^{(1)} \psi_k^{(0)}.
\end{eqnarray}
%We find out that the normalization of the perturbed wavefunction is clearly greater than unity:
%\begin{eqnarray}
%\braket{\psi_n|\psi_n}&=&\braket{\psi_n^{(0)}+\sum_k a_{kn}^{(1)\ast} \psi_k^{\ast(0)}|\psi_n^{(0)}+\sum_m a_{mn}^{(1)} \psi_m^{(0)}} \nonumber \\
%&=&1+0+0+\sum_k|a_{kn}^{(1)}|^2 > 1.
%\end{eqnarray}
Let us further analyze our result for a two level system. The perturbed wavefunction for the ground state $E_1$ is given by,
\begin{eqnarray}
\psi_1^{(\lambda)}&=&(1+\lambda_1 U_{11})\psi_1^{(0)}+\lambda_2 W_{12} \psi_2^{(0)}.
\end{eqnarray}
Here we suppose that $a_{11}^{(1)}=\lambda_1 U_{11},\,a^{(1)}_{12}=\lambda_2 W_{12}$. Let us calculate the inner product which is satisfied to yield the fidelity susceptibility, 
Finally we suggest the following expression for the fidelity susceptibility $\chi_F$ for a time-driving system
\begin{widetext}
\begin{align}
\chi_{ij}=\Big[\frac{\braket{\psi_1(\lambda)|\partial_{\lambda_i }\psi_1(\lambda)}}{\braket{\psi_1|\psi_1}}\Big]
\Big[\frac{\braket{\psi_1(\lambda)|\partial_{\lambda_j }\psi_1(\lambda)}}{\braket{\psi_1|\psi_1}}\Big]
+2 \frac{\braket{\psi_1(\lambda)|\partial_{\lambda_i }\partial_{\lambda_j }\psi_1(\lambda)}}{\braket{\psi_1|\psi_1}}\delta_{ij}.\label{chiF}
\end{align}
\end{widetext}
Note that $d\hat{s}^2=\chi_{ij}\delta\lambda_i\delta\lambda_j$ defines a 
Riemannian metric on a manifold ${\cal M}$ which is a family of (positive definite) inner products -- for all differentiable vector fields $\lambda_1,\lambda_2$ on ${\cal M}$, that defines a smooth function ${\cal M}\rightarrow R^2$ on coordinate space $(\lambda_i)^{2}$.  An explicit form for the metric can be written as follows:
\begin{equation}
ds^2=\chi_{11}d\lambda_1^2+2{\rm Re}(\chi_{12})d\lambda_1 d\lambda_2 + \chi_{22}d\lambda_2^2,
\end{equation}
or its equivalent form,
\begin{equation}
ds^2=\chi_{ij}{(t)}d\lambda_i d\lambda_j.
\end{equation}
%It is adequate to consider it as co-dimension one spacelike slicing of the bulk spacetime:
%\begin{equation}
%ds^2_{2+1}=\chi_{ij}{(t)}d\lambda_i d\lambda_j-dt^2.
%\end{equation}
%and define a co-dimension two minimal surface via minimization of the following action
%\begin{equation}
%\mbox{Area}=\int{\chi_{ij}{(t)}d\lambda_i d\lambda_j}.
%\end{equation}

%%%%%%%%%%%%%%%%%%%%%%%%%%%%%%%%%%%%%%%%%%%%%%%
\section{Fidelity susceptibility in the Landau-Zener problem}
\label{IV}
%%%%%%%%%%%%%%%%%%%%%%%%%%%%%%%%%%%%%%%%%%%%%%%
In the previous section we introduced a general formulation for fidelity susceptibility for time deriving potential. In this section, we will investigate a concrete example, inspired from Landau-Lifshitz cookbooks \cite{Landau}. The system under consideration is a two-level quantum system initially prepared in ground state. The model named as Landau-Zener problem. The aim is to calculate $\chi_F$ matrix using (\ref{chiF}). 
The ground state is defined by $n=0$ and it satisfies:
\begin{equation}
\braket{H_0}_{0}\quad\leq \quad\braket{H_0}_{\rm{Excited\,state}}.
\end{equation}
The energy levels for the unperturbed Hamiltonian $H_0$ is defined as  
 $E_a={E_1,\,E_2}$ and it is convenient to define a frequency basis for the system,
\begin{equation}
\omega_{12}=\frac{E_2-E_1}{\hbar} > 0.
\end{equation}
As a two-level system, $E_1=E_0=E_{min}$, consequently we have:
\begin{equation}
E_2>E_1.
\end{equation}
The following two total wavefunctions of a two-level system $E_2>E_1$ are defined using the orthogonality realization:
\begin{equation}
\Psi_1=\sum_k a_{k1}(t) \psi_k^{(0)}, \ \ 
\Psi_2=\sum_k a_{k2}(t) \psi_k^{(0)},
\end{equation}
where
\begin{equation}
a_{kn}=\delta_{kn}-\frac{i}{\hbar}\int{V_{kn}e^{i\omega_{kn}t}dt}.
\label{akn}
\end{equation}
Next we propose a specific form of the potential as
\begin{equation}
V=F e^{-i\omega t}+G e^{i\omega t},
\end{equation}
where $F$ and $G$ are time-independent operators. If $V_{nm}=V_{mn}^\ast$ then we obtain $G_{nm}=F_{mn}^\ast$. In this situation, the matrix element takes the form, 
\begin{eqnarray}
V_{kn}(t) &=& V_{kn} e^{i\omega_{kn}t}
          =F_{kn} e^{i(w_{kn}-\omega)t} + F_{kn}^\ast e^{i(w_{kn}+\omega)t}.
\label{vkn}
\end{eqnarray}
Substituting (\ref{vkn}) into (\ref{akn}) and performing an integration, we obtain
\begin{equation}
a_{kn}^{(1)}=-\frac{F_{kn}e^{i(\omega_{kn}-w)t}}{\hbar(\omega_{kn}-\omega)}-\frac{F_{kn}^\ast e^{i(\omega_{kn}+w)t}}{\hbar(\omega_{kn}+\omega)},
\end{equation}
where we have assumed that $\omega_{kn} \neq \pm \omega$. Note that the matrix element for an arbitrary operator $O$  is given by:
\begin{eqnarray}
O_{mn}(t)&=&O_{mn}^{(0)} e^{i\omega_{nmt}}+O_{mn}^{(1)}(t),
\end{eqnarray}
where
\begin{widetext}
\begin{equation}
O_{mn}^{(1)}(t)=e^{i\omega_{nm}t}\Big(\sum_k{\Big[\frac{O_{nk}^{(0)}F_{km}}{\hbar(\omega_{km}-\omega)}+\frac{O_{km}^{(0)}F_{nk}}{\hbar(\omega_{kn}+\omega)}\Big]e^{-i\omega t}+\Big[\frac{O_{nk}^{(0)}F_{mk}^\ast}{\hbar(\omega_{mk}+\omega)}+\frac{O_{km}^{(0)}F_{kn}^\ast}{\hbar(\omega_{nk}-\omega)}\Big]e^{i\omega t}}\Big).
\end{equation}
\end{widetext}
To be more concrete when choosing $O=H$ and  $H_{nk}^{(0)}=E_k \delta_{nk}$, the matrix form for $H$ in zeroth order reads,
\begin{widetext}
\begin{eqnarray}
H_{nm} &=& E_m \delta_{nm} e^{i \omega_{nm}t}\nonumber\\&&
-e^{i\omega_{nm}t}\Big(\sum_k{\Big[\frac{E_k \delta_{nk} F_{km}}{\hbar(\omega_{km}-\omega)}+\frac{E_k \delta_{mk} F_{nk}}{\hbar(\omega_{kn}+\omega)}\Big]e^{-i\omega t}+\Big[\frac{E_k \delta_{nk} F_{mk}^\ast}{\hbar(\omega_{mk}+\omega)}+\frac{E_k \delta_{mk} F_{kn}^\ast}{\hbar(\omega_{nk}-\omega)}\Big]e^{i\omega t}}\Big).
\end{eqnarray}
\end{widetext}
If $F$ is real, i.e., $F_{mn}=F_{nm}^\ast$, we obtain the following expression for a matrix representation of $H$ up to the first-order perturbation,
\begin{widetext}
\begin{eqnarray}
H_{nm} &=&E_n \delta_{nm} e^{i \omega_{nm}t}- e^{i \omega_{nm}t} F_{nm}\omega_{nm} \left(\frac{e^{-i \omega t}}{\omega_{nm}-\omega}+\frac{e^{i \omega t}}{\omega_{nm}+\omega}\right).
\end{eqnarray}
\end{widetext}
Note that the diagonal elements are commonly parametrized by 
$H_{nn}=E_n$ and the off diagonal ones are 
\begin{equation}
H_{n \neq m} = -e^{i \omega_{nm}t} F_{nm}\omega_{nm}   \left(\frac{e^{-i \omega t}}{\omega_{nm}-\omega}+\frac{e^{i \omega t}}{\omega_{nm}+\omega}\right).
\end{equation}
%\begin{equation}
%$H_{12}=H_{21}^\ast=\begin{pmatrix} E_1 & \omega_0 e^{-i \omega_0 t} F_{12} \times (\frac{e^{i \omega t}}{\omega-\omega_0}-\frac{e^{-i \omega t}}{\omega+\omega_0}) \\ -\omega_0 e^{-i \omega_0 t} F_{12} \times (\frac{e^{i \omega t}}{\omega+\omega_0}-\frac{e^{-i \omega t}}{\omega-\omega_0}) & E_2 \end{pmatrix}$ \\
%\end{equation}
\par
For the two-level system, it is still plausible to obtain
\begin{eqnarray}
\nonumber
H_{12}&=&(H_{21})^\ast=
\omega_0 F_{12} \left (\frac{e^{i (\omega-\omega_0) t}}{\omega-\omega_0}-\frac{e^{-i (\omega+\omega_0) t}}{\omega+\omega_0}\right).\label{H12}
\end{eqnarray}
The wavefunction coefficients read as follows:
\begin{equation}
a_{11}^{(1)}=i\frac{F_{11}}{\hbar \omega}\sin(\omega t),
\end{equation}
and
\begin{eqnarray}
a_{21}^{(1)}
&=&-\frac{F_{12}^\ast}{\hbar} \Big[\frac{e^{-i\omega t}}{\omega_0-\omega}+\frac{e^{i\omega t}}{\omega_0+\omega}\Big] .
\end{eqnarray}
Therefore, the total perturbed wavefunction for the ground state is given by,
\begin{equation}
\Psi_1=\left(\frac{F_{11}}{\hbar \omega}\sin(\omega t)\right)\psi_1^{(0)}-\frac{F_{12}^\ast}{\hbar}\left(\frac{e^{-i\omega t}}{\omega_0-\omega}+\frac{e^{i\omega t}}{\omega_0+\omega}\right)\psi_2^{(0)}.
\end{equation}
It is reasonable to parametrize perturbed matrix elements as follows:
\begin{eqnarray}
F_{11}&=&\lambda_1 V_{11} \\
F_{12}&=&\lambda_2 W_{12}.
\end{eqnarray}
In terms of these parameters, we obtain
\begin{widetext}
\begin{equation}
\Psi_1=\left(\frac{i V_{11}}{\hbar \omega} \sin\omega t\right) \lambda_{1} \psi_1^{(0)}  -  \frac{i W_{12}^\ast}{\hbar}e^{i \omega_0 t}\left(\frac{e^{-i\omega t}}{\omega_0-\omega}+\frac{e^{i\omega t}}{\omega_0+\omega}\right)\lambda_2 \psi_2^{(0)}.\label{wavefunction0}
\end{equation}
\end{widetext}
By defining two auxiliary functions,
\begin{eqnarray}
\alpha(t)&=&\frac{i V_{11}}{\hbar \omega} \sin\omega t, \\
\beta(t)&=&\frac{-i W_{12}^\ast}{\hbar}\left(\frac{e^{i(\omega_0-\omega)t}}{\omega_0-\omega}+\frac{e^{i(\omega_0+\omega)t}}{\omega_0+\omega}\right),
\end{eqnarray}
and using (\ref{chiF}), we end up with the matrix elements for $\chi_F$ as follows:
\begin{eqnarray}
\chi_{11}&=&\frac{1}{2 \lambda_1}\frac{1}{1+|\frac{\beta }{\alpha}|^2(\frac{\lambda_2}{\lambda_1})^2},\\
\chi_{12}&=&\frac{1}{2\lambda_1}\frac{1+|\frac{\beta }{\alpha}|^2 (\frac{\lambda_2}{\lambda_1})}{1+|\frac{\beta }{\alpha}|^2 (\frac{\lambda_2}{\lambda_1})^2},\\
\chi_{22}&=&\frac{1}{2\lambda_1}\frac{|\frac{\beta }{\alpha}|^2 }{1+|\frac{\beta }{\alpha}|^2 (\frac{\lambda_2}{\lambda_1})^2}.
\end{eqnarray}
Note that here $\lambda_2\neq \lambda_1$ to have the non singular metric $\chi_{ij}$. In our model, 
\begin{eqnarray}
\nonumber\abs{\frac{\beta}{\alpha}}^2=|\frac{2\omega W_{12}^\ast}{ V_{11}}|^2\Big(\frac{\omega ^2 \cos ^2\left( \omega _0 t\right)+\omega _0^2 \sin ^2\left( \omega
   _0 t\right) \cot ^2( \omega t)}{\left(\omega ^2-\omega _0^2\right){}^2}
\Big)
\end{eqnarray}
We are interested in high frequencies where $\omega\gg\omega_0$. In this case we have
\begin{eqnarray}
\abs{\frac{\beta}{\alpha}}^2\approx \abs{\frac{4 W_{12}^\ast}{ V_{11}}}^2  \cos ^2\left( \omega _0 t\right).
\end{eqnarray}
Finally, by defining $\gamma\equiv |\frac{4 W_{12}^\ast}{ V_{11}}|^2>0 $, we have the following approximated form for fidelity susceptibility at high frequencies and ultraviolet (UV) regime as follows:
\begin{eqnarray}
\chi_{11}&=&\frac{1}{2 \lambda_1}\frac{1}{1+\gamma  \cos ^2\left( \omega _0 t\right)(\frac{\lambda_2}{\lambda_1})^2},\\
\chi_{12}&=&\frac{1}{2\lambda_1}\frac{1+\gamma  \cos ^2\left( \omega _0 t\right) (\frac{\lambda_2}{\lambda_1})}{1+\gamma  \cos ^2\left( \omega _0 t\right)(\frac{\lambda_2}{\lambda_1})^2},\\
\chi_{22}&=&\frac{1}{2\lambda_1}\frac{\gamma  \cos ^2\left( \omega _0 t\right) }{1+\gamma  \cos ^2\left( \omega _0 t\right) (\frac{\lambda_2}{\lambda_1})^2}.
\end{eqnarray}
The information metric, measures the distance between two quantum states close to each other in UV regime and is given as follows:
\begin{widetext}
\begin{eqnarray}
&&ds^2=\frac{1}{2 \lambda_1(1+\gamma  \cos ^2\left( \omega _0 t\right)(\frac{\lambda_2}{\lambda_1})^2)}\Big[d\lambda_1^2
+2\big(1+\gamma  \cos ^2\left( \omega _0 t\right) (\frac{\lambda_2}{\lambda_1})\big)d\lambda_1d\lambda_2+\gamma  \cos ^2\left( \omega _0 t\right)d\lambda_2^2
\Big].
\end{eqnarray}
\end{widetext}
This metric could be dual to a non relativistic time dependent bulk theory via  Maldacena's AdS/CFT correspondence \cite{Maldacena} in a same methodology as presented in \cite{prl}.   

%%%%%%%%%%%%%%%%%%%%%%%%%%%%%%%%%%%%%%%%%%%%%%%%%
\section{Measurement $\chi_F$ using quantum noise setups }
\label{V}
In recent years, the time-dependent systems phase transitions have been investigated in references \cite{report} -\cite{2017arXiv170105152B}.
%%%%%%%%%%%%%%%%%%%%%%%%%%%%%%%%%%%%
In Ref.\cite{report}, the universal scaling behavior in a one-dimensional quantum Ising model subject to time-dependent sinusoidal modulation in time of its transverse magnetic field has been illustrated. This scaling behavior existed in various quantities, e.g. concurrence, entanglement entropy, magnetic and  fidelity susceptibility. Based on an Ising spin chain and with periodically varying external magnetic field along the transverse direction the authors, in Ref.\cite{2017arXiv171107769M}, investigated the microscopic quantum correlations dynamics of the bipartite entanglement and quantum discord. 

\begin{figure}
	\includegraphics[width=8.0cm]{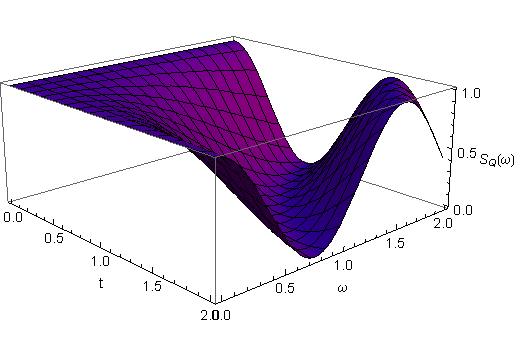}
	\caption{The plot shows the noise function $S_{Q}(\omega)$ as functions of time ($t$) and frequency ($\omega$).} \label{noise}
\end{figure}

%%%%%%%%%%%%%%%%%%%%%%%%%%
In this section, we mainly focus on frequency spectrum of the quantum system resulting from the  quantum noise function. Let us assume a generalized Hamiltonian $H=H_0+\lambda V$ with $\lambda$ denoting the control parameter. The quantum noise spectrum of the driven Hamiltonian $V$ can be defined as
\begin{eqnarray}
S_Q(\omega)&=&\sum_{n\neq0}\left|\langle\phi_n|V|\phi_0\rangle\right|^2\delta(\omega-E_n+E_0),\label{noisefunction}
\end{eqnarray}
where $\ket{\phi_n}$ is the eigenstate of the Hamiltonian $H(\lambda)$ and we assumed $E_n$ as non-degenerate energy levels of the whole system. Note that the quantum noise function $S_Q(\omega)$ can be constructed from the excited states $E_n>E_0$. In our model, the ground state wavefunction is given in Eq.(\ref{wavefunction0}). Here we can rewrite the noise function (\ref{noisefunction}) using matrix element given in Eq.(\ref{vkn}) as follows:
\begin{eqnarray}
&&S_Q(\omega)=
=2|\lambda_2 W_{12}|^2\cos^2\omega t.
\label{noisefunction2}
\end{eqnarray}
We plot the noise function $S_{Q}(\omega)$ versus time ($t$) and frequency ($\omega$)  illustrated in Fig.(\ref{noise}). 

As well known, the fidelity susceptibility plays an important role in QPTs stemming from the fact that it is always possible to describe the universality classes of QPTs without specifying the type of the symmetry of the system. However, it is adequate to ask whether we can measure $\chi_F$ using  experimental setups. It has been shown that recently the $\xi_F$ is related to the quantum noise spectrum of the time-driven Hamiltonian \cite{You2015}. It is remarkable to relate $\xi_F$ to $S_Q(\omega)$ using Kronig-Penney transformation:
\begin{eqnarray}
\chi_F&=&\int_{-\infty}^{\infty}d\omega\frac{S_Q(\omega)}{\omega^2}.
\end{eqnarray}
Bearing in  mind that the following definition of derivative of any analytic function $f(z)$ provides a useful tool:
\begin{eqnarray}
f(z)&:&\mathcal{C}\to\mathcal{C},\nonumber\\f^{(n)}(z)&=&\frac{\Gamma(n+1)}{2\pi i}\int \frac{f(w)}{(w-z)^{n+1}}dw,\label{complex}
\end{eqnarray}
where $\Gamma(n+1)=\int_{0}^{\infty}e^{-t}t^{n}dt $ is a Gamma function. Using  (\ref{complex}) we clearly observe that \cite{You2015}:
\begin{eqnarray}
\chi_F=\pi i\frac{d^2 S_Q(\omega)}{d\omega^2}|_{\omega=0}. 
\end{eqnarray}
It is clearly stated that $ S_Q(\omega)$ can be measured in laboratory, see Ref.\cite{fd1966}. Consequently we verify that the $\chi_f$ could be measured in the laboratory, as well. Particularly the Landau-Lifshitz model with $\chi_F$ presented in Eqs.(54)-(56) provides a useful machinery to study the universal scaling behavior of $\chi_F$.
%%%%%%%%%%%%%%%%%%%%%%%%%%%

%%%%%%%%%%%%%%%%%%%%%%%%%%%%%%%%%%%%%%%%%%%%%%%%%%
\section{ $\mathcal{O}{(\delta\lambda)^3 }$ Missing term }
\label{VI}
%%%%%%%%%%%%%%%%%%%%%%%%%%%%%%%%%%%%%%%%%%%%%%%%%%
In this section, we highlight higher order corrections up to the $\mathcal{O}(\delta\lambda^3)$ of $\chi_F$. 
 It is noteworthy to figure out higher order terms, i.e., 
 the coefficient of  $\delta
\lambda^2$  using the expressions  given above. Remember that 
\begin{eqnarray}
\ket{\psi(\lambda+\delta\lambda)}=\ket{\psi(\lambda)}+\delta\lambda \partial_{\lambda }\ket{\psi(\lambda)}+\frac{\delta\lambda^2}{2}\partial^2_{\lambda }\ket{\psi(\lambda)}.
\end{eqnarray}
Let us compute the following inner product:
\begin{widetext}
\begin{eqnarray}
\braket{\psi(\lambda)|\psi(\lambda+\delta\lambda)}\approx\braket{\psi(\lambda)|\psi(\lambda)}+\delta\lambda \braket{\psi(\lambda)|\partial_{\lambda }\psi(\lambda)}+\delta\lambda^2\braket{\psi(\lambda)|\partial^2_{\lambda }\psi(\lambda)}+...\,\,,
\end{eqnarray}
\end{widetext}
where the ellipses denote higher order (correction) terms. Consequently, we obtain the following expressionfor the third-order fidelity susceptibility as follows:
\begin{eqnarray}
&&\zeta_F=\frac{\braket{\psi(\lambda)|\partial_{\lambda }\psi(\lambda)}\braket{\psi(\lambda)|\partial^2_{\lambda }\psi(\lambda)}}{|\braket{\psi|\psi}|^2}.
\end{eqnarray}
%Now using Eqs.(\ref{78}-\ref{80}) we obtain:
%\begin{eqnarray}
%\braket{\psi|\partial^2_{\lambda}\psi}&=&2\lambda c(a(t)+\lambda b+\lambda^2 c)+|W(t)|^{-2}\Big[\frac{i\hbar}{\lambda}(\dot a+\lambda \dot b+\lambda^2 \dot c)-V(t)(a+\lambda b+\lambda^2 c)
%\Big]\times\\&&\nonumber\times\Big[2 i\hbar \lambda^{-3}\dot a-2 V(t)c\Big].
%\end{eqnarray}
The above equation defines a higher order correction to the usual fidelity susceptibility. The corresponding metric is a Finsler manifold in which the general information metric is characterized by the following form:
\begin{eqnarray}
ds^2=\chi_{ij}d\lambda_id\lambda_j+(\zeta_{ijk}d\lambda_id\lambda_j\lambda_k)^{\frac{2}{3}}+...\,\,.
\end{eqnarray}
It is worth noting that the distant between two quantum states in any quantum theory can be quantified not only by fidelity but also with higher order cubic quantity defined by $\zeta_F$. We note here that the corresponding tensor form for $\zeta_F$ is given by:
\begin{eqnarray}
&&(\zeta_F)_{ijk}=\frac{\braket{\psi(\lambda)|\partial_{\lambda_i }\psi(\lambda)}\braket{\psi(\lambda)|\partial_{\lambda_j }\partial_{\lambda_k}\psi(\lambda)}}{|\braket{\psi|\psi}|^2}.
\end{eqnarray}
It will be very interesting to find bulk dual for this new tensor in a similar way recently suggested for fidelity susceptibility as a maximal volume in the AdS spacetime \cite{prl}.

%%%%%%%%%%%%%%%%%%%%%%%%%%%%
\section{Summary}

In this work, we have presented a simple and straightforward approach to compute distance between quantum states responsible for the fidelity susceptibility in quantum model for a time-dependent system describing a two-level atom coupled to a time-driven external field. Analytically we have investigated the behavior of fidelity susceptibility in the time-driven quantum model in which the potential $V$ is time-dependent. Interestingly, the information metric induced by fidelity susceptibility can be nicely achieved. We also plotted the obtained noise function and found that the noise function eternally lasts long in our model. We have also derived for the first time a higher-order rank-3 tensor as third-order fidelity susceptibility for having a model beyond fidelity susceptibility.  

It will be very interesting to find bulk dual for this new tensor in a similar way recently suggested for fidelity susceptibility as a maximal volume in the AdS spacetime \cite{prl}. Moreover, as mentioned in Refs.\cite{Gan:2017qkz,Alishahiha:2017cuk}, our understanding of quantum gravity may be satisfied using quantum information theory along with holography. This may allow us to further examine a possible connection between the fidelity susceptibility and holographic complexity and may shed new light on the deeper understanding of quantum gravity.

%%%%%%%%%%%%%%%%%%%%%%%%%%%%%%%%%%%
\section{Acknowledgment} 
D. Momeni and M. Al Ajmi would like to acknowledge the support of Sultan Qaboos University under the Internal Grant (IG/SCI/PHYS/19/02).
%%%%%%%%%%%%%%%%%%%%%%%%%%%%%%%

\end{document}